\documentclass[utf8]{FrontiersinVancouver} 

\setcitestyle{square} 
\usepackage{hyperref,lineno,microtype,subcaption}
\usepackage[onehalfspacing]{setspace}

\def\keyFont{\fontsize{8}{11}\helveticabold }
\def\firstAuthorLast{N Jia {et~al.}} 
\def\Authors{Ning Jia\,$^{1}$, Xing-Dong Zhao\,$^{2}$, Wen-Rong Qi\,$^{2}$ and Jing Qian\,$^{3,*}$}

\def\Address{$^{1}$The Public Experimental Center, University of Shanghai for Science and Technology, Shanghai 200093, China \\
$^{2}$School of Physics, Henan Normal University, Xinxiang 453000, China \\
$^{3}$State Key Laboratory of Precision Spectroscopy, Quantum Institute for Light and Atoms, Department of Physics, School of Physics and Electronic Science, East China Normal University, Shanghai 200062, China}
\def\corrAuthor{Jing Qian}
\def\corrEmail{jqian1982@gmail.com}

\begin{document}
\onecolumn
\firstpage{1}

\title[ ]{Ultraprecise off-axis atom localization with hybrid fields} 

\author[\firstAuthorLast ]{\Authors} 
\address{} 
\correspondance{} 

\extraAuth{}

\maketitle

\begin{abstract}

\section{}
Atom localization enables a high-precision imaging of the atomic position, which has provided vast applications in fundamental and applied science. In the present work, we propose a scheme for realizing two-dimensional off-axis atom localization in a three-level $\Lambda$ system. Benefiting from the use of a hybrid coupling field which consists of one Gaussian beam and one Laguerre-Gaussian beam,
our scheme shows that the atoms can be localized at arbitrary position with a high spatial resolution. Taking account of realistic experimental parameters, our numerical simulation predicts that the atoms can be precisely localized with a spatial resolution of $\sim 200$ nm in the range of 
a radial distance of a few micrometers to the beam core. Our results provide a more flexible way to localize atoms in a two-dimensional system, possibly paving one-step closer to the nanometer-scale atom lithography and ultraprecise microscopy.

\tiny
 \keyFont{ \section{Keywords:} off-axis localization, Laguerre-Gaussian beam, ultraprecise, quantum interference} 
\end{abstract}

\section{Introduction}


Nowadays the Laguerre-Gaussian(LG) beam \cite{Babiker2018} has engendered tremendous advanced applications\cite{tweezer,tw_NP,tw1,nlo1996,nlo2017,optics}.
For example, it is widely used in the superresolution fluorescence microscopy such as the stimulated emission depletion \cite{sted1,sted2} and minimal photon fluxes \cite{minflux1,minflux2} in order to overcome the diffraction limit. 
Another approach to this target is utilizing the spatially-dependent coherent light-matter interaction in atom-light coupling systems \cite{p4expr,expr1,expr2} which essentially depends on a spatially modulated atom-light coupling. Via the detection of spontaneously-emitted photon \cite{emit,p1a,emit1,emit2},
level population \cite{fluorescence,p1,pop_ad1,p2,gamma,p3a,p3} absorption \cite{absorb1,absorb2,absorb3} and gain \cite{gain,gain2}, subwavelength-scale atom localization can be obtained. 

As far as we know, atom localization with LG beams can exhibit a large number of advantages \cite{LG1,LG2}. For example, LG beam has a doughnut intensity spot naturally, which may avoid the need of two orthogonal standing-wave(SW) fields for generating spatially modulated atom-light coupling in a two-dimensional(2D) atom localization system. That fact largely reduces the complexity of experimental implementation.
Moreover, it is easier to create single excitation spot in its core by a LG beam. In traditional SW-based localization schemes, due to the periodicity of the SW field intensity there may exist more than one localization spots within single optical wavelength. Therefore after one-time measurement, the probability of finding atoms at a certain position can be deeply reduced. So far, some approaches have been proposed to break this periodicity of SW fields, via utilizing {\it e.g.} the sensitivity of light-matter interactions to the light phase in a closed-loop atomic system \cite{loop2011absorb,loop2016}, the interference of multiple SW fields with different wavelengths and phases \cite{kakb,kakb2}. These methods, however, will increase the complexity of experimental setup. 
Although LG beam has the above advantages in localization, it can only localize atoms in the vicinity of its beam core where the laser intensity is close to zero. An off-axis atom localization must be accompanied by the movement of the LG beam itself which adds to extra complexity.


In traditional SW localization schemes, the superposition of multiple SW lasers with different wavelengths and phases is commonly adopted for reaching a single excitation point \cite{kakb,kakb2}. Besides this effect between two LG beams can show interesting patterns such as optical Ferris wheels where the light intensity can be modulated to be zero in certain positions \cite{FW}. Inspired by these contributions, in the present work, we study a 2D off-axis atom localization in a three-level $\Lambda$ system in which a Gaussian beam serves as the probe field and a LG beam together with a Gaussian beam as the hybrid coupling field. 
The quantum interference effect between these two beams(LG and Gaussian) can achieve a unique zero-intensity spot at arbitrary position.
We show that, by appropriately tuning the ratio of peak amplitudes between the LG and any Gaussian beams, atoms can be localized at arbitrary position with a certain distance to the beam core. Both the spatial resolution and radial distance of localization can be flexibly manipulated via tuning laser Rabi frequencies. Depending on the numerical simulation with experimental parameters our scheme enables the realization of an efficient off-axis 2D atom localization, accompanied by a best spatial resolution $\sim 200$ nm and a radial distance of a few micrometers. Our scheme provides a more convenient route to the target of ultraprecise off-axis 2D atom localization.


\section{Theoretical strategy}

\begin{figure}
\centering
\includegraphics[width=0.45\textwidth]{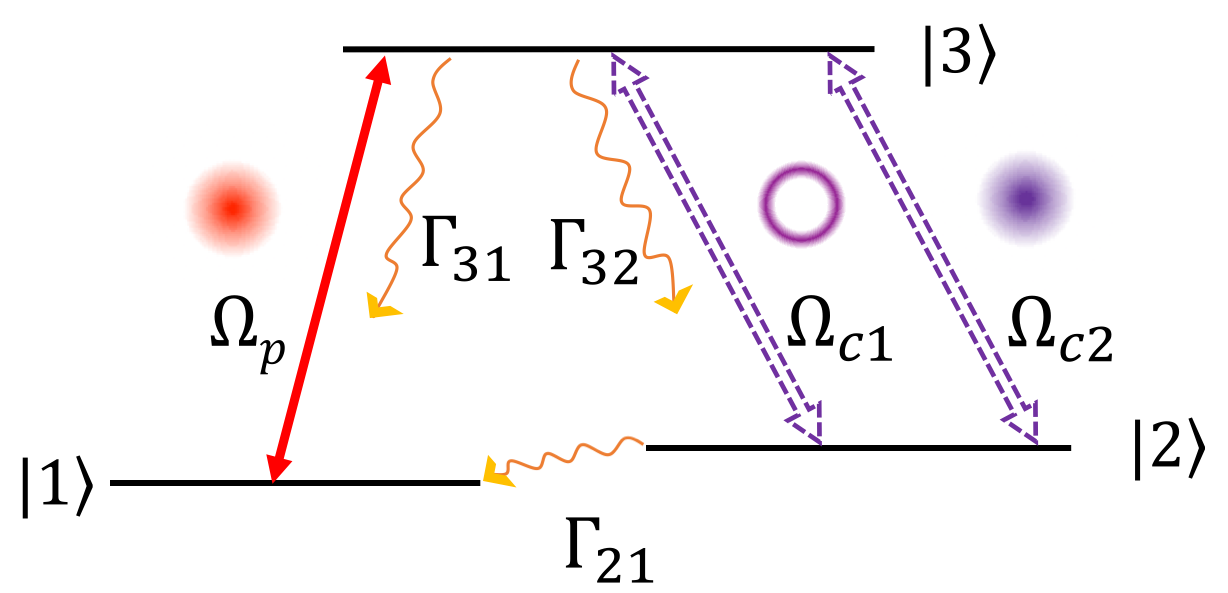}
\caption{Schematic of a $\Lambda$-type three-level system where the probe field $\Omega_p$ for the transition of states $|1\rangle$ and $|3\rangle$ is a Gaussian beam. The coupling field between $|3\rangle$ and $|2\rangle$ is composed by one LG beam $\Omega_{c1}$ and one Gaussian beam $\Omega_{c2}$. $\Gamma_{nm}$ denotes the spontaneous decay rates from $|n\rangle$ to $|m\rangle$.}
\label{fig_model}
\end{figure}

To describe the scheme mechanism we consider a simple three-level $\Lambda$ system as displayed in Fig.\ref{fig_model}, where states $\left|1\right\rangle$ and $\left|3\right\rangle$ are resonantly coupled by a weak probe field $ \Omega_p$ and states $\left|3\right\rangle$ and $\left|2\right\rangle$ are connected by another coupling field $\Omega_{c2}$ with a zero detuning.
In order to realize an off-axis atom excitation, we have assumed that the probe and coupling beams as Gaussian beams, which are
 \begin{equation}
 \Omega_{i}(r)=\Omega_{i0}e^{-\frac{r^2}{W_{i}^2}}
 \label{Omgp}
 \end{equation}
with $i=p,c2$ and $\Omega_{i0}$ the peak amplitude and $W_i$ the Gaussian spot size. Remarkably, states $\left|3\right\rangle$ and $\left|2\right\rangle$ is also coupled by a second LG field $\Omega_{c1}$ at the same time, as \cite{LG_func}
 \begin{equation}
 \Omega_{c1}(r,\theta)=\Omega_{c10}(\frac{r}{W_{c1}})^{|l|}e^{-\frac{r^2}{W_{c1}^2}}e^{il(\theta+\theta_{c1})}  
 \label{Omgc}
 \end{equation}
where $\Omega_{c10}$, $W_{c1}$, $\theta_{c1}$ and $l$ are the peak amplitude, the beam waist, the initial phase and the winding number, respectively. $r$ and $\theta$ are the cylindrical radius and the azimuthal angle.
Here, we take the winding number $l=1$ which enables  a single-spot excitation. Because other higher-order modes with $l>1$ would lower the localization precision by redistributing the atomic population among multiple azimuthal nodes.
To our knowledge this three-level model can be experimentally realized by the D1 line of ultracold $^{87}$Rb atoms with energy levels $|1\rangle=\left|5S_{1/2},F=1\right\rangle$, $|2\rangle=\left|5S_{1/2},F=2\right\rangle$, and $|3\rangle=\left|5P_{1/2},F=2\right\rangle$. Based on Ref. \cite{expr2}, we assume the decay rates from $|3\rangle\to|1\rangle$ and $|3\rangle\to|2\rangle$ are equal, typically calculated by $\Gamma_{31}=\Gamma_{32}=2\pi\times5.75$ MHz. The decay rate between two hyperfine ground states $|1\rangle$ and $|2\rangle$ is $\Gamma_{21}=5$ kHz, satisfying $\Gamma_{21}\ll\Gamma_{31},\Gamma_{32}$ \cite{gamma1} so the lifetime of $\left|2\right\rangle$ is about $200\mu$s. 
The beam width is $W_i=W_{c1}=W$ estimated to be same for simplicity.
Under the frozen-gas limit where the atomic center-of-mass keeps unvaried we can take a measurement for the population on state $|2\rangle$ by collecting its fluorescence signals with a CCD camera and a well-localized position distribution of atoms could facilitate this measurement \cite{expr2}.



Taking account of a frozen atomic gas the time evolution of the systematic density-matrix elements can be described by ($\hbar=1$) \cite{evolution_func}
\begin{eqnarray}
\label{Bloch_eqs}
\dot{\rho}_{11} &=& \Gamma_{31}\rho_{33}+\Gamma_{21}\rho_{22}-i\Omega_{p}\left(\rho_{31}-\rho_{13}\right),\nonumber\\
\dot{\rho}_{33} &=& -\left(\Gamma_{31}+\Gamma_{32}\right)\rho_{33}-i\Omega_{p}\left(\rho_{13}-\rho_{31}\right)-i\Omega_{c}\left(\rho_{23}-\rho_{32}\right) ,\nonumber\\
\dot{\rho}_{12} &=&  -\gamma_{12}\rho_{12}-i\Omega_{p}\rho_{32}+i\Omega_{c}\rho_{13}, \label{motion} \\
\dot{\rho}_{13} &=& -\gamma_{13}\rho_{13}-i\Omega_{p}\left(\rho_{33}-\rho_{11}\right)+i\Omega_{c}\rho_{12} ,\nonumber\\
\dot{\rho}_{23} &=& -\gamma_{23}\rho_{23}+i\Omega_{p}\rho_{21}-i\Omega_{c}\left(\rho_{33}-\rho_{22}\right),\nonumber
 \end{eqnarray}
with $\rho_{nm}=\rho_{mn}^{*}$ and $\sum_{n=1}^{3}\rho_{nn}=1$ means the conservation. The population on state $|2\rangle$ is solved by $\rho_{22}=1-\rho_{11}-\rho_{33}$. In deriving Eq.(\ref{motion}) we have defined
\begin{equation}
  \Omega_c=\Omega_{c1}+\Omega_{c2}
\label{Oc}  
\end{equation}
representing the superposition of two coupling fields. $\Gamma_{nm}$ denotes the spontaneous decay from $\left|n\right\rangle$ to $\left|m\right\rangle$ and $\gamma_{nm}$ is defined as
\begin{equation}
 \gamma_{12}=\Gamma_{21}/2, \gamma_{13}=(\Gamma_{31}+\Gamma_{32})/2, \gamma_{23}=(\Gamma_{32}+\Gamma_{31}+\Gamma_{21})/2.
\end{equation}
The steady solutions of Eq.(\ref{Bloch_eqs}) can be obtained by assuming
$\dot{\rho}_{nm}=\dot{\rho}_{nn}=0$. Due to the presence of decay $\Gamma_{21}$, it is intuitive that $\rho_{22}$ decreases with $\Gamma_{21}$. Luckily, accounting for the condition of $\Gamma_{21}\ll\Gamma_{31(32)}$ that makes the effect of $\Gamma_{21}$ negligible \cite{gamma,gamma1}, $\rho_{22}$ takes a simple form of
\begin{equation}
  \rho_{22}(r,\theta)=\frac{1}{1+I_c(r,\theta)/I_p(r)}  
  \label{popu}
\end{equation}
where $\Gamma_{31}=\Gamma_{32}$, $\Gamma_{21}=0$ are used. $I_c = \left|\Omega_{c1}+\Omega_{c2} \right|^2$, $I_p = \left|\Omega_{p} \right|^2$ stand for the laser intensities. Note that $\rho_{22}(r,\theta)$ reveals  position-dependent feature due to the use of several structured fields. From Eq.(\ref{popu}), it is apparent that the condition $I_c(r_{loc},\theta)\ll I_p(r_{loc})$ will cause $\rho_{22}\rightarrow1$ which means a perfect atomic confinement can be achieved at arbitrary position $r_{loc}$ in our scheme.

\section{Off-axis localization}
\label{offaxis}

According to Eq.(\ref{Oc}) together with the definitions in Eqs.(\ref{Omgp}-\ref{Omgc}), the intensity of the hybrid coupling field can be written as 
\begin{equation}
 I_{c}(r,\theta)=|\Omega_{c1}+\Omega_{c2}|^2=\frac{\Omega_{c10}}{W^{2}}e^{-\frac{2r^2}{W^{2}}}\left|r+\kappa_{c}We^{-i(\theta+\theta_{c1})}\right|^2
 \label{Ic}
 \end{equation}
where the peak ratio is $\kappa_{c}=\Omega_{c20}/\Omega_{c10}$ which can be tuned by 
$\Omega_{c20}$ if $\Omega_{c10}$ is fixed. Note that this hybrid coupling field is composed by one LG beam and one Gaussian beam which resonantly couple states $|2\rangle$ and $|3\rangle$ at the same time.
Finally we can arrive at an analytical solution to the equation $I_{c}(r,\theta)=0$, {\it i.e.} the perfect condition of localization can be reached at
\begin{equation}
  (r_{loc},\theta_{loc})=(\kappa_cW,\pi-\theta_{c1}) 
  \label{position}
\end{equation}
where the population $\rho_{22}$ attains 1.0 in principle. That means atoms can be precisely placed at any desired position $(r_{loc},\theta_{loc})$ with a very high probability. While in fact, owing to the influence from intrinsic noises in the experimental setup, the observed localization resolution is quite limited. In Section \ref{sec_Fes} we will discuss the fluctuation of laser intensities, the steady time as well as the noise from atomic thermal motion, in order to present a practical estimation for the experimental observation.
Besides we have to point out that, benefiting from the interference between two hybrid coupling fields $\Omega_{c1}$ and $\Omega_{c2}$ \cite{Qiu2019}, the
localization position  $(r_{loc},\theta_{loc})$ can be widely adjusted by the beam parameters which is not restricted merely at the beam core as in most previous works \cite{LG1,LG2}.

 \begin{figure}
\centering
\includegraphics[width=0.5
\textwidth]{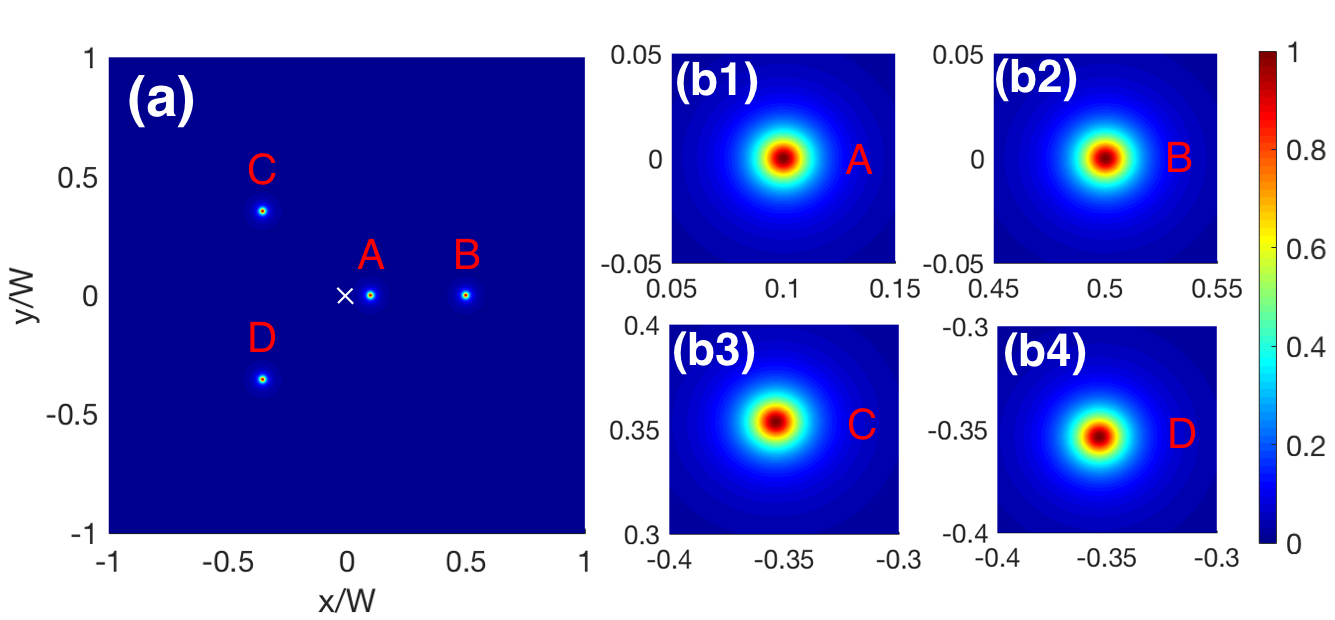}
\caption{Off-axis atom localization. (a) Off-axis localized positions A, B, C and D with respect to $(\kappa_c, \theta_{c1})=(0.1,\pi), (0.5,\pi), (0.5,\pi/4), (0.5,-\pi/4)$. The white cross denotes the center of light beams. (b1)-(b4) show the amplified images for the localized atomic positions which have an off-axis feature. Here $\Omega_{p0}/\Omega_{c10}=0.01$.}
\label{fig_direction}
\end{figure}

As illustrated in Fig. \ref{fig_direction}a, we show that atoms denoted as the steady population $\rho_{22}$ on state $|2\rangle$, can be confined in any spatial position $(r_{loc},\theta_{loc})$ by changing the parameters $(\kappa_c, \theta_{c1})$. For example when $(\kappa_c, \theta_{c1})=(0.1,\pi), (0.5,\pi), (0.5,\pi/4), (0.5,-\pi/4)$,
Fig. \ref{fig_direction}a explicitly shows the off-axis atom localization at different positions as labeled by A$\sim$D. Fig. \ref{fig_direction}(b1-b4) amplify the distribution of atoms at different localized places. It is apparent that the spatial resolution of atom localization keeps unchanged for different $\kappa_c$ and $\theta_{c1}$.
 Therefore, thanks to the zero intensity point($I_c(r,\theta)=0$) created by the interference between two light beams $\Omega_{c10}$ and $\Omega_{c20}$, our scheme can realize an effective off-axis localization at arbitrary position in a 2D space.

\section{Ultra-precise localization}
\label{sec_precision}

\begin{figure}[h]
\centering
\includegraphics[width=0.5\textwidth]{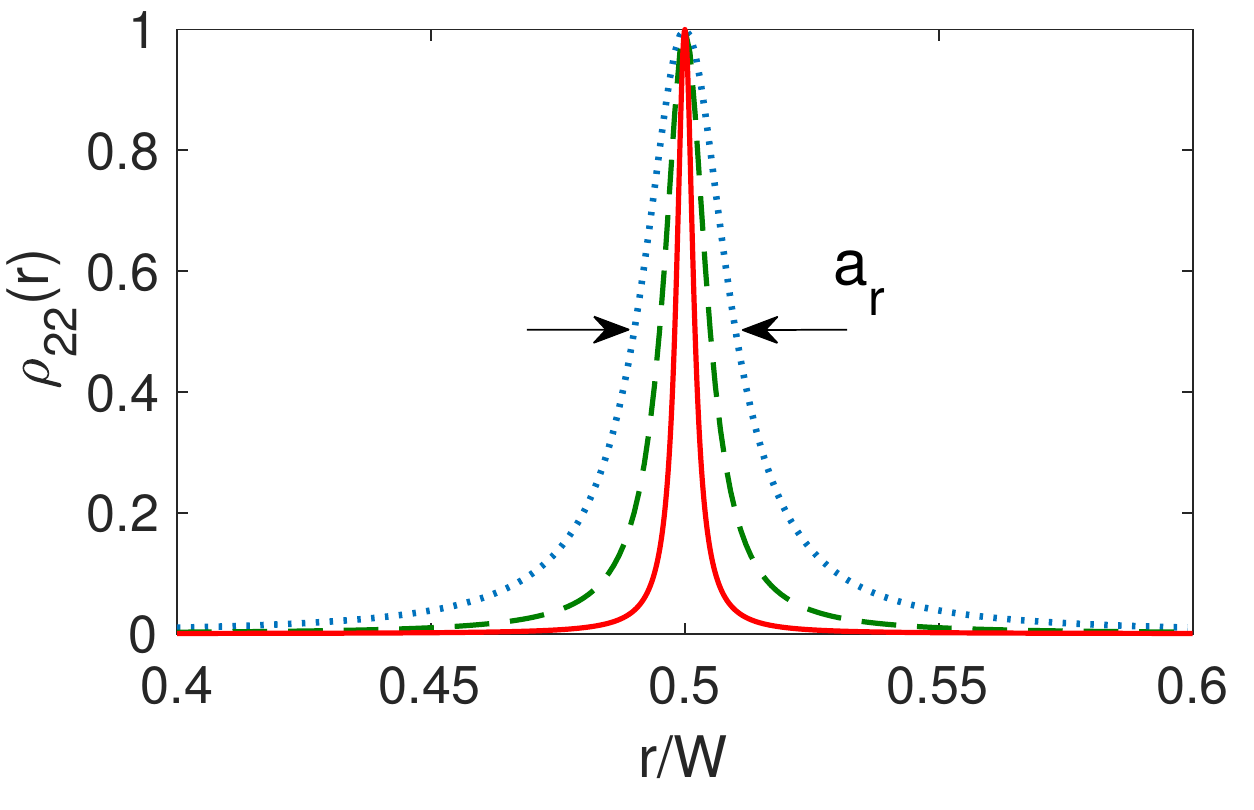}
\caption{Steady population distribution $\rho_{22}(r)$ along the radial direction $r$ for $\kappa_p = \Omega_{p0}/\Omega_{c10}=$ 0.01 (blue-dotted), 0.005 (green-dashed) and 0.002 (red-solid). $a_r$ is the FWHM which characterizes the spatial resolution of localization. Here we use $\kappa_c=0.5$.}
\label{ar}
\end{figure}

The quality of localization also depends on a high spatial resolution which is characterized by the full width at half maximum(FWHM) of the steady distribution $\rho_{22}(r,\theta)$. A narrower linewidth indicates that the position of atoms can be well-resolved within a smaller range. By replacing the profiles of light fields[see Eqs.(\ref{Omgp}) and (\ref{Omgc})] the expression of $\rho_{22}(r)$ takes an explicitly Lorentz form
\begin{equation}
    \rho_{22}(r)=\frac{1}{1+\frac{(r-\kappa_cW)^2}{\kappa_p^2W^2}}
\label{rho22_Lorentz}
\end{equation}
in which we have omitted the azimuth angle by letting $\theta=\pi-\theta_{c1}$ and paid attention to the variation of $\rho_{22}(r)$ along the radial direction. We treat the FWHM of function $\rho_{22}(r)$ as a measurement to localization, which can also be analytically solved,
\begin{equation}
    a_r=2\kappa_pW.
\label{arr}
\end{equation}

In Fig. \ref{ar} we plot the steady distribution $\rho_{22}(r)$ {\it vs} $r$ for different peak ratios $\kappa_p$. Clearly a weaker probe field leads to the atomic population more confined in the vicinity of localization point $r=r_{loc}=0.5W$, promising for a higher-resolution localization. For example we find that $a_r=0.02W$ when $\kappa_p=0.01$; but this value is decreased by one order of magnitude which is $a_r=0.004W$ as $\kappa_p$ reduces to 0.002. From Eq. (\ref{arr}), it is intuitive that $a_r\rightarrow0$ if $\kappa_p\ll1$ enabling an ultra-precise localization under a sufficiently weak probe field.
However a realistic system arises the fact that the time for a steady state becomes much longer in the weak-probe limit, resulting in the atomic motion non-negligible.
We will discuss this point in section \ref{sec_Ts}. 

\section{Feasibility discussion}
\label{sec_Fes}
The numbers presented in this work are considered from $^{87}$Rb where the lifetime of state $|3\rangle$($=|5P_{1/2},F=2\rangle$) is 27.7ns \cite{Volz_1996} leading to the decay rates $\Gamma_{31}=\Gamma_{32}=2\pi\times5.75$ MHz, and the lifetime of $|2\rangle$($=|5S_{1/2},F=2\rangle$) is 200 $\mu$s arising the decay rate to be $\Gamma_{21}=5$ kHz. We assume that the co-propagating probe and coupling lasers are overlapping in space and have a same beam width $W=5$ $\mu$m.
As explicitly presented in sections \ref{offaxis} and \ref{sec_precision},
our scheme can achieve an ultraprecise off-axis atom localization due to the flexible manipulation of peak ratios $\kappa_c$ and $\kappa_p$, 
together with the azimuth angle $\theta_{c1}$.
Due to the rotational invariance we ignore $\theta_{c1}$ by focusing on the radial distance $r$.
However, as for an experimental implementation these parameters are also restrained. In this section, we numerically solve the spatial resolution $a_r$ and the peak value of $\rho_{22}(r)$ by evolving the motional equations (\ref{motion}) with more realistic conditions coming from measurement.

\subsection {Laser intensity noise}

\begin{figure}
\centering
\includegraphics[width=0.6\textwidth]{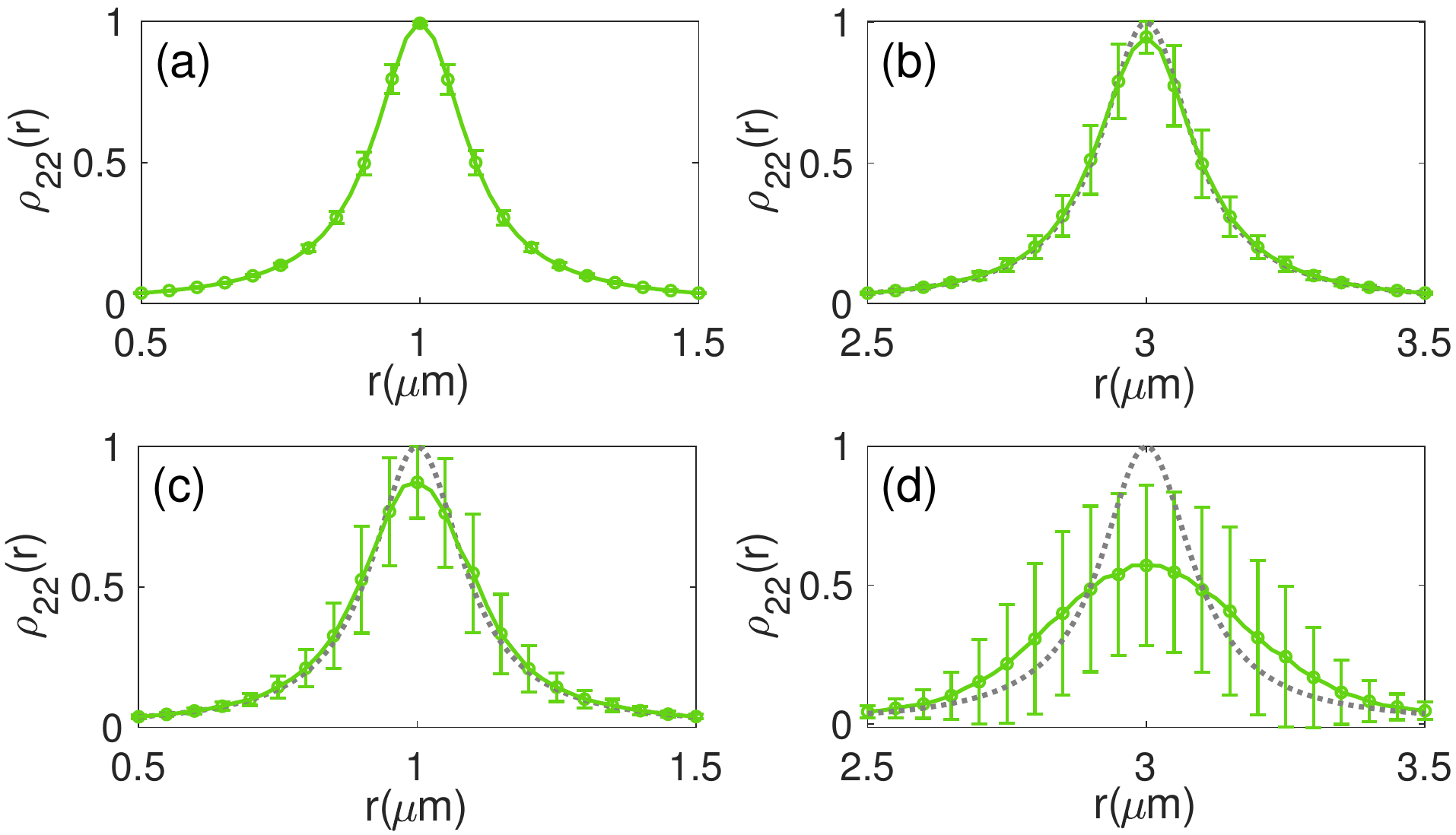}
\caption{Radial population distribution $\rho_{22}(r)$ under different intensity noises which are given by (a-b) $\xi=1.0\%$ and (c-d) $\xi=5.0\%$ at different positions.
We take 500 measurements for each point denoted by the error bar and the average result is shown by the green solid line. For comparison the black-dotted line indicates the result without any intensity noise {\it i.e.} $\xi=0$. Here $\Omega_{c20}/2\pi = (30,90) $ MHz, respectively for (a),(c) and (b),(d), corresponding to the localization positions $r_{loc}=(1.0,3.0)$ $\mu$m.
Other parameters are $\Omega_{p0}/2\pi = 3$ MHz, $\Omega_{c10}/2\pi = 150$ MHz and $W=5$ $\mu$m. 
}
\label{IntensityNoise}
\end{figure}

To obtain realistic results evaluating experimental conditions we introduce a perturbed laser intensity by adding a random intensity noise $\delta\Omega_i$($i=p,c1,c2$) to the peak value $\Omega_{i0}$\cite{intensity_noise2009,intensity_noise2018}. The resulting fluctuated Rabi frequencies $\Omega_{i0}^{\prime}$ can be written as 
\begin{equation}
    \Omega_{i0}^{\prime}=\Omega_{i0}+\delta\Omega_i.
\end{equation}
In the calculation we assume $\delta\Omega_i/\Omega_{i0}\in[-\xi,\xi]$ and pay attention to the radial population distribution $\rho_{22}(r)$. During each measurement, the perturbation term $\delta\Omega_i$ can be a random number obtained from the range of $[-\xi,\xi]\Omega_{i0}$. By taking account of sufficient measurements, the average result can show the realistic observation in experimental setup. Notice that a larger Rabi frequency leads to stronger laser noise since $\delta\Omega_i\propto\Omega_{i0}$.

Figure \ref{IntensityNoise} illustrates the distribution of steady population $\rho_{22}(r)$ under the influence of laser intensity noise which is characterized by the factor $\xi$. By comparing Fig. \ref{IntensityNoise}(a-b) and (c-d) it is apparent that a bigger $\xi$ will give rise to a broadened population distribution with smaller peak values, which lowers the precision of localization. Furthermore, as for atoms localized closer to the beam core($r=0$) the intensity noise $\delta\Omega_{c2}$[$\propto \Omega_{c20}$] is smaller due to $r_{loc}=\kappa_c W$. Therefore by positioning atoms far from the beam core the observation will suffer from a stronger laser intensity noise, in turn yielding a lower-quality localization, 
see Fig. \ref{IntensityNoise}(a,c) and (b,d).
This fact gives a limitation to our protocol that the atoms can not be placed very far from the beam core. A rough estimation(not shown) shows that the average peak value of $\rho_{22}(r)$ will be smaller than 0.2 if the radial localization distance is $r_{loc} >10 \mu$m. In experiment a well control for the laser intensity noise can improve the scheme performance.

\subsection {Time needed for a steady state}
\label{sec_Ts}

From section \ref{sec_precision}, we have known that an ultra-precise localization with $a_r\to 0$ in principle relies on a sufficiently small $\kappa_p$, {\it i.e.} $\Omega_{p0}\ll\Omega_{c10}$. While this condition leads to the time $T_s$ for reaching steady localization much longer. Because $T_s$ is inversely proportional to the exact laser Rabi frequencies. For a longer $T_s$, the atomic thermal motion does play roles and the frozen-gas approximation fails.
A discussion for the effect of atomic thermal motion can be seen in Sec.\ref{therm}.
An efficient localization requires that $T_s$ is so short as to make the atomic movement during the steady time negligible. In the calculation we consider atoms under the temperature $T=1$ $\mu$K \cite{expr2} with a most probable velocity $v_p=\sqrt{2k_B T/M}\approx1.4$ cm/s, where $k_B$ is the Boltzmann constant and $M$ is the atomic mass. We introduce a new constraint to the resolution factor $a_r$
\begin{equation}
    v_p T_{s}^{max} \leq a_r/10
    \label{lim}
\end{equation}
which means the real time $T_s$ for a steady state should be smaller than $T_{s}^{max}$ so as to make the atomic motion negligible during the measurement.

\begin{figure}
\centering
\includegraphics[width=0.5\textwidth]{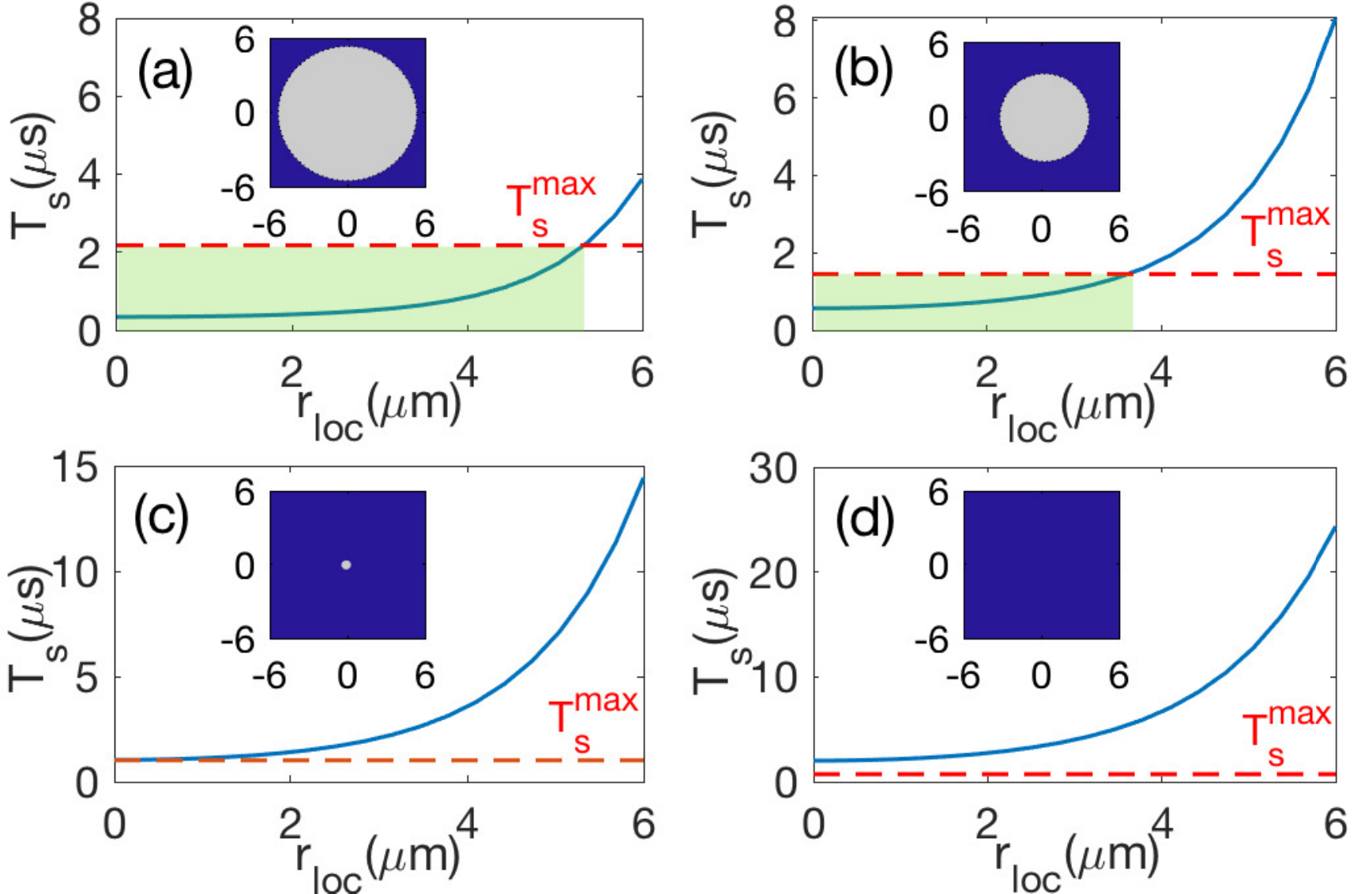}
\caption{Steady time $T_s$ {\it vs} the radius distance $r_{loc}$ under (a) $\Omega_{p0}/2\pi = 4.5$ MHz and $a_r=300$ nm, (b) $\Omega_{p0}/2\pi = 3.0$ MHz and $a_r=200$ nm, (c) $\Omega_{p0}/2\pi = 2.1$ MHz and $a_r=141$ nm, (d) $\Omega_{p0}/2\pi = 1.5$ MHz and $a_r=100$ nm. The red-dashed line denotes the maximal $T_s$ permitted for an efficient localization. The shaded-green region stands for the radial range where atoms can be localized. Insets: Effective off-axis localization is enabled within the gray disk. Here $\Omega_{c10}/2\pi=150$ MHz, $W=5$ $\mu$m, $\Omega_{c20}=r_{loc} \Omega_{c10}/W$, $T=1$ $\mu$K and $\Gamma_{21}=5$ kHz.}
\label{Ts}
\end{figure}

Figure \ref{Ts} exhibits the steady time $T_s$ as a function of the localization distance $r_{loc}$ for different peak probe Rabi frequencies $\Omega_{p0}$. Here $T_s$ is obtained by numerically evolving the master equation (\ref{Bloch_eqs}), taking account of all spontaneous decays. From (a) to (d), as decreasing $\Omega_{p0}$ we find that the steady time $T_s$(blue-solid) increases significantly; although the position of atoms can be well-resolved with a better spatial resolution($a_r$ becomes smaller) at the same time. According to the constraint (\ref{lim}) the maximal steady time $T_{s}^{max}$ permitted for localization is labeled by the red-dashed line in the figure. When $T_s<T_s^{max}$ atoms can obtain a robust localization. Obviously, in (a) and (b) where the spatial resolution $a_r$ is relatively large, atoms can be well localized within a wider radial range $r_{loc}<5.3$ $\mu$m and $r_{loc}<3.7$ $\mu$m. Insets explicitly show the area of off-axis localization which is denoted as a gray disk. In fact via an appropriate adjustment of $\kappa_c$ and $\theta_{c1}$, atoms can be confined at arbitrary position inside the gray disk.

Whereas, when $\Omega_{p0}$ is reduced to $2\pi\times 2.1$ MHz[see Fig. \ref{Ts}c] the reduction of $a_r$ causes $T_{s}^{max}\leq T_s$ persistently. In this case only atom positioned at the beam core can be accurately confined so the protocol of off-axis localization fails. Furthermore, if $a_r < 141$ nm {\it e.g.} $a_r =100$ nm as in (d), the steady time $T_s$ keeps larger than the required $T_s^{max}$ calculated by Eq. (\ref{lim}) so no atoms could be localized. Because during such a longer steady time $T_s$ most atoms have been moved away from the localization spot caused by their thermal motions, leading to a poor resolution (also see the discussion in Sec.\ref{therm}).
Therefore, based on our analysis, we treat $a_r = 141$ nm as the best spatial resolution yet atoms can only be localized at the beam core. An effective off-axis localization needs to be at the expense of the spatial resolution. For example, a resolution of $a_r = 200$ nm(300 nm) can be obtained within a localized radius of $r_{loc}<3.7$ $\mu$m(5.3 $\mu$m), see the insets of Fig.\ref{Ts}(a-b) for a more visible representation. In addition since the steady time is inversely proportional to the exact Rabi frequencies the limitation for a best off-axis localization can further be overcome by a stronger coupling laser. {\it e.g.} when $\Omega_{c10}/2\pi=300$ MHz and $\Omega_{p0}/2\pi=2.7$ MHz the best spatial resolution of our protocol can be reduced to 91 nm if atoms are localized in the beam core(not shown).

\subsection{Noise from atomic thermal motion}\label{therm}

\begin{figure}
\centering
\includegraphics[width=0.8\textwidth]{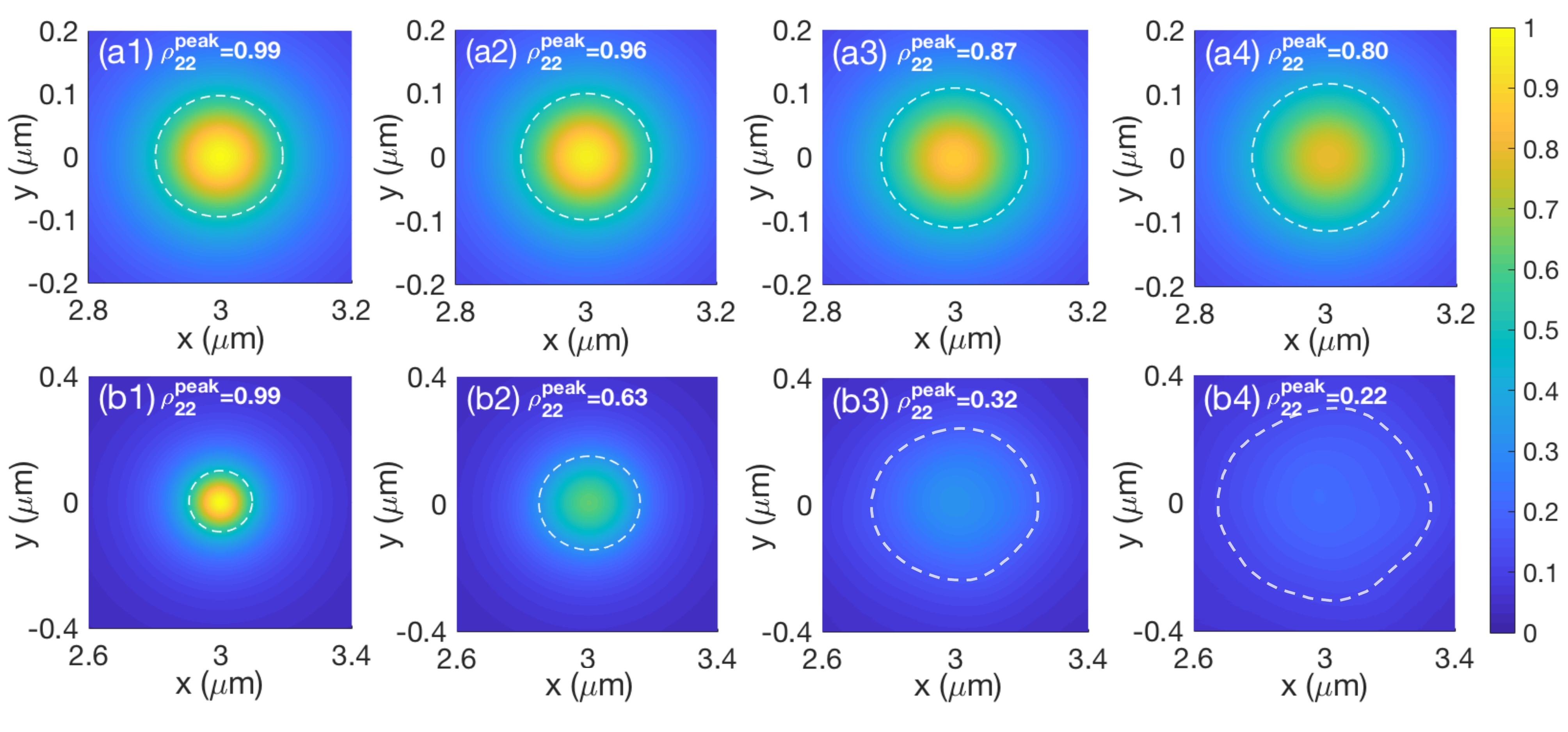}

\caption{(a1-a4) The 2D population distribution of $\rho_{22}(r)$ under different temperatures $T=(0,1,5,10)\mu$K. The peak value of $\rho_{22}(r)$ is given in the picture and the diameter of white-dashed rings stands for the spatial resolution $a_r$ which is $a_r=(200, 206, 222, 247)$ nm, respectively. Here we assume the measurement time is $T_{meas}=1$ $\mu$s. Analogous to (a1-a4), (b1-b4) show the case of $T_{meas}=5$ $\mu$s and the calculated resolution is $a_r=(200, 308, 530, 687)$ nm. Every point is obtained by averaging over 500 measurements. }
\label{movement}
\end{figure}

In a real experimental setup due to atomic thermal motion, the laser intensity 'seen' by atoms would have a strong perturbation which inversely
brings a noise on detecting the steady atomic population. Here we consider atoms move randomly in space whose velocities satisfy a two-dimensional Maxwell–Boltzmann distribution\cite{Huo2022}
\begin{equation}
 f(v_x,v_y)=\frac{1}{\pi v_{p}^2}e^{-(v_x^2+v_y^2)/v_{p}^2}.
\end{equation}
with $v_p$ the most probable velocity defined by $v_p=\sqrt{2k_B T/M}$. Other interatomic collisions are ignored here.
During the $j$th measurement we assume a simple uniform motion of atoms by letting
\begin{equation}
  (x_j,y_j)\rightarrow(x_j+v_x T_{meas}, y_j+v_y T_{meas}) 
  \label{xy}
\end{equation}
where $(v_x,v_y)$ are obtained stochastically from the velocity function $f(v_x,v_y)$ and $T_{meas}$ is the time for single measurement. By inserting Eq.\ref{xy} into Eqs.(\ref{Omgp}) and (\ref{Omgc}) atoms can feel a fluctuated Rabi frequency $\Omega_i(t)$ ($i=p,c1,c2$) for each measurement. The final results are based on an average of 500 times random samplings of the velocity $(v_x,v_y)$.

In figure \ref{movement} we show the calculated population distribution $\rho_{22}(r)$ under sufficient measurements in the x-y frame. Clearly, from (a1) to (a4) due to a larger probable velocity of atoms caused by the growing temperature, the peak value $\rho_{22}^{peak}$ has an explicit decrease together with a lower spatial resolution $a_r$. For example when $T=1$ $\mu$K, $a_r=206$ nm which is close to the value at $T=0$. Because the average distance of atoms during each measurement $T_{meas}=1$ $\mu$s is only $v_pT_{meas}\approx14$ nm which is much smaller than $a_r$. However, as for a higher temperature the movement of atoms during each measurement can cause a bigger effect making the precision of atom localization poor. See the case of $T=10$ $\mu$K in (a4) we observe that $\rho_{22}^{peak}=0.8$ and $a_r=247$ nm. For comparison in (b1-b4) we also study the case of a longer measurement time($T_{meas}=5$ $\mu$s) where atoms can move farther, leading to a very poor spatial resolution at a finite temperature. We numerically show that, at $T=10$ $\mu$K the distribution of atomic population $\rho_{22}(r)$ has become slightly deformed with its peak value(spatial resolution) as low as $\rho_{22}^{peak}=0.22$($a_r=687$ nm). That fact means such a long-time measurement has made most atoms away from the localization spot via their thermal movements.
Therefore
a faster measurement accompanied by a lower environment temperature can facilitate a high-quality atom localization.

\section{Conclusions}

To conclude, our scheme presents a novel 2D atom localization having both ultraprecise and off-axis features. Differing from the previous works using a single LG field we adopt
a LG beam together with a Gaussian beam as the hybrid coupling field. The previous contributions can only localize atom in the beam core where the intensity of coupling field is zero. While our protocol shows that atoms can be localized at arbitrary position due to the effect of quantum interference between these two coupling beams that leads to a zero-intensity spot in space. Our numerical simulation confirms that with an appropriate adjustment for the peak ratios of laser Rabi frequencies a wider off-axis localization range as well as a higher-quality spatial resolution can be achieved at the same time.
Under experimentally-feasible parameters an estimation for the implementation of realistic off-axis atom localization is predicted, promising for a resolution of $\sim 200$ nm and a localized radius of a few $\mu$m. In addition we also discuss the weakness of our scheme when some intrinsic quantum noises from imperfect measurement, including laser intensity noise, limited steady time and atomic thermal motion, are considered.
Our approach may provide unique application to atomic lithography with more flexibility and better resolution \cite{li}. An extension to the 
3D off-axis atom localization is possible by implementing a spatial modulation to the probe detuning which is our next-step work \cite{LG2}.

\section*{Conflict of Interest Statement}

The authors declare that the research was conducted in the absence of any commercial or financial relationships that could be construed as a potential conflict of interest.

\section*{Author Contributions}

The idea was first conceived by NJ. NJ was responsible for the physical modeling, the numerical calculations, and writing the original draft under the supervision of JQ. JQ contributed to review and editing, JQ verified results of the theoretical calculation,  X-DZ contributed to editing the draft. X-DZ and W-RQ contributed to the discussion of the results.

\section*{Funding}
This work was supported by the National Natural Science Foundation of China under Grants No. 12104308, No. 12174106, No. 11474094, No.11104076 and No. 12104135; by the Science and Technology Commission of Shanghai Municipality under Grant No.18ZR1412800.

\section*{Data Availability Statement}
The original contributions presented in the study are included in the article/supplementary material; further inquiries can be directed to the corresponding authors.

\bibliographystyle{Frontiers-Vancouver} 

\bibliography{reference_LG}

\end{document}